\documentclass[12pt,preprint]{aastex}

\newcommand{\etal}{{et al}.}            

\usepackage{longtable}
\usepackage{lscape}

\begin{document}

\title{The XBo{\"o}tes Chandra Survey \\Paper II: \\ 
 The X-ray Source Catalog}

\author{  Almus Kenter\altaffilmark{1},Stephen S. Murray\altaffilmark{1}, 
William R. Forman\altaffilmark{1}, Christine Jones\altaffilmark{1},
\\
Paul Green\altaffilmark{1}, Christopher S. Kochanek\altaffilmark{5},
Alexey Vikhlinin\altaffilmark{1}, Daniel Fabricant\altaffilmark{1},
\\
 Giovani Fazio\altaffilmark{1}, Katherine Brand\altaffilmark{2},
Michael J. I. Brown\altaffilmark{2}\altaffilmark{,6}, Arjun Dey\altaffilmark{2},
Buell T. Jannuzi\altaffilmark{2}, \\
Joan Najita\altaffilmark{2}, Brian McNamara\altaffilmark{3}, Joseph
Shields\altaffilmark{3}, \\
and Marcia Rieke\altaffilmark{4}}

\begin{abstract}

 We present results from a Chandra survey of
 the nine square degree Bo{\"o}tes field of the NOAO Deep Wide-Field Survey (NDWFS).  
This XBo{\"o}tes  survey consists of 126 separate
contiguous ACIS-I observations each of approximately 5000 seconds in duration.
These  unique Chandra observations allow us  
to search for large scale structure and to  calculate  X-ray source statistics over a 
wide, contiguous field of view with arcsecond angular resolution
and  uniform coverage. Optical spectroscopic follow-up observations
and the rich NDWFS data set will allow us to identify
 and classify these X-ray selected sources.

Using wavelet decomposition, we detect 4642 point sources
with n~$\ge$~2 counts. In order to keep our
detections $\sim$~99\% reliable, we limit our list to 
sources with n~$\ge$~4 counts. 
For a 5000 second observation and assuming a canonical unabsorbed AGN type
x-ray spectrum, a 4 count on-axis source corresponds to a 
 flux  of $  4.7 \times 10^{-15}$~ergs~cm$^{-2}$~s$^{-1}$
 in the soft (0.5--2~keV) band,
$ 1.5 \times 10^{-14}$~ergs~cm$^{-2}$~s$^{-1}$ in the hard (2--7~keV)
band and $ 7.8 \times 10^{-15}$~ergs~cm$^{-2}$~s$^{-1}$ in the full (0.5--7~keV) band.
The full 0.5--7~keV band n~$\ge$~ 4 count list has 3293 point sources.
In addition to  the  point sources,  43  extended  sources
have been detected consistent, with the depth of these observations
and the number counts of clusters. We present here the X-ray catalog for
the XBo{\"o}tes survey, including source positions, X-ray fluxes, 
hardness ratios and their uncertainties.
 We  calculate and present the  differential  number of sources per  flux density interval,
$N(S)$, for the point sources.  
In the soft (0.5--2~keV) band, $N(S)$ is well fit by a broken power-law  with slope of $ 2.60^{+0.11}_{-0.12}$
 at bright
fluxes and  $1.74^{+0.28}_{-0.22}$ for faint fluxes.
The hard source $N(S)$ is well described by a single power-law with an index of  
$-2.93^{+0.09}_{-0.09}$

\end{abstract}
\dataset {ADS/Sa.CXO\#def/XBootes}

\altaffiltext{1}{Harvard-Smithsonian Center for Astrophysics, 60
  Garden St., Cambridge, MA 02138}

\altaffiltext{2}{National Optical Astronomy Observatory, 950 N. Cherry
  Ave.,Tucson, Az 85719}

\altaffiltext{3}{Ohio University Department of Physics and Astronomy,
  Athens, OH 45701}

\altaffiltext{4}{Steward Observatory, University of Arizona, 933
  N. Cherry Ave., Tucson, AZ 85751}

\altaffiltext{5}{The Ohio State University, Department of Astronomy,
  Columbus, OH 43210}

\altaffiltext{6}{Department of Astrophysical Sciences, Princeton
  University, Peyton Hall, Ivy Lane, Princeton, NY 08544}

\section{Introduction}

The NOAO Deep Wide-Field Survey (NDWFS) is 
 a deep optical and near-infrared wide-field
 imaging survey that
is designed to investigate the existence and evolution of large scale structures at
 redshifts z~$>$1.
The NDWFS consists of two $\sim$ 9 degree square regions; one in the constellation
Cetus and one in the constellation Bo{\"o}tes. The regions have been chosen to 
maximize the depth and amount of multiwavelength coverage.
The northern Bo{\"o}tes field has been observed  
by the Gemini Observatory, Spitzer, the VLA,  Chandra and  Hubble.
Details of the NDWFS program are presented in \citet{Buell99} and references
therein. Here we present the source list for the XBo{\"o}tes Chandra
Survey \citep{SSM1}. 

Detecting optical counterparts for 
weak X-ray sources from previous X-ray
missions has proven
to be problematic; the error circles on
weak X-ray source positions typically were large and the 
optical counterparts for many of the
 X-ray sources
 are  optically faint  \citep[$R>24$][]{Has98}.
Deep optical observations  within the error circles
of the  X-ray positions often presented several 
possible optical counterparts, increasing the probability
for incorrect identifications
\citep{Boyle,Georg,McHard}.

Unlike earlier wide area surveys, the  XBo{\"o}tes data have arcsecond  resolution
and
broad  energy response to 10 keV; this allows the survey to sample unique and highly
absorbed sources and to provide a source list for
  nearly un-ambiguous optical and  spectroscopic follow-up observations.
The angular resolution, uniform coverage and large contiguous field of view allow us to 
search for evidence of  structure on both small and large angular scales,
relatively free of edge effects and position inaccuracy biases.
The survey was described and the distribution of the X-ray selected sources was
 examined for evidence of
Large Scale Structure in a companion paper \citep{SSM1}, hereafter Paper~I.
Optical matching and identification of these X-ray selected sources  
is discussed in a second companion paper \citep{KateB}. 
Multi-fiber spectrometer spectra using the MMT Hectospec  as part of the
 AGN and Galaxy Evolution Survey (AGES)
 are being used to determine
redshifts for these X-ray selected sources \citep{Chris_K}. Photometric redshifts
will be determined for those sources not observable by Hectospec.
The resulting three dimensional distribution will be
examined for evidence of spatial correlation in a
series of forthcoming papers. 
 
This paper addresses the following aspects
of the Chandra analysis: the data set and preliminary
data processing; the source
detection algorithms;  the source catalog; 
flux and hardness ratios for the point sources;
 and the number of sources per flux density interval.
The complete version of the source catalog is in the electronic edition of
the Journal.  The printed edition contains only a sample.
A machine readable version of the table is also available.
\section{The Chandra Observations}
The 126 pointings comprising this  survey are centered on 
$\alpha_{J2000} \approx $ $14$ hours $32$ minutes
and
 $\delta_{J2000} \approx $ $34^{\circ}06'$ and were completed during March and April of 2003.
 At that time,  the ACIS instrument was  at $-120^{\circ}$~C.
In addition to the four ACIS-I chips, (\verb+ccd_id=0,1,2,3+), an
ACIS-S chip (\verb+ccd_id=6+) was also active. The ACIS-S data is far off the telescope axis and
has not yet been analyzed.
These observations were taken and processed in Very Faint Mode (VFM) \citep{VFM}.
VFM reduces the background by 
a factor of 2.5 near 0.5 keV, 1.4 near 1 keV, 1.1 between 2--5~keV, 
and 1.4--1.5 above 7 keV; this improves sensitivity, especially
to faint diffuse extended objects such as clusters of galaxies.
The data underwent the standard Chandra event processing
and, in addition, the data were ``de-streaked'' and all afterglow events were
removed using CIAO data processing tools  \citep{CXC_proc}. 
Data were examined for intervals of excess background flaring with 
the \verb+lc_clean+ tool described in \citet[and references therein]{MAXIMbkg}.
Six observations had background rates approximately two
times greater than the average rate. Nevertheless, these
higher background rate observations do not suffer
from any significant loss of sensitivity for point source
detection.

\begin{figure}[p]
\plotone{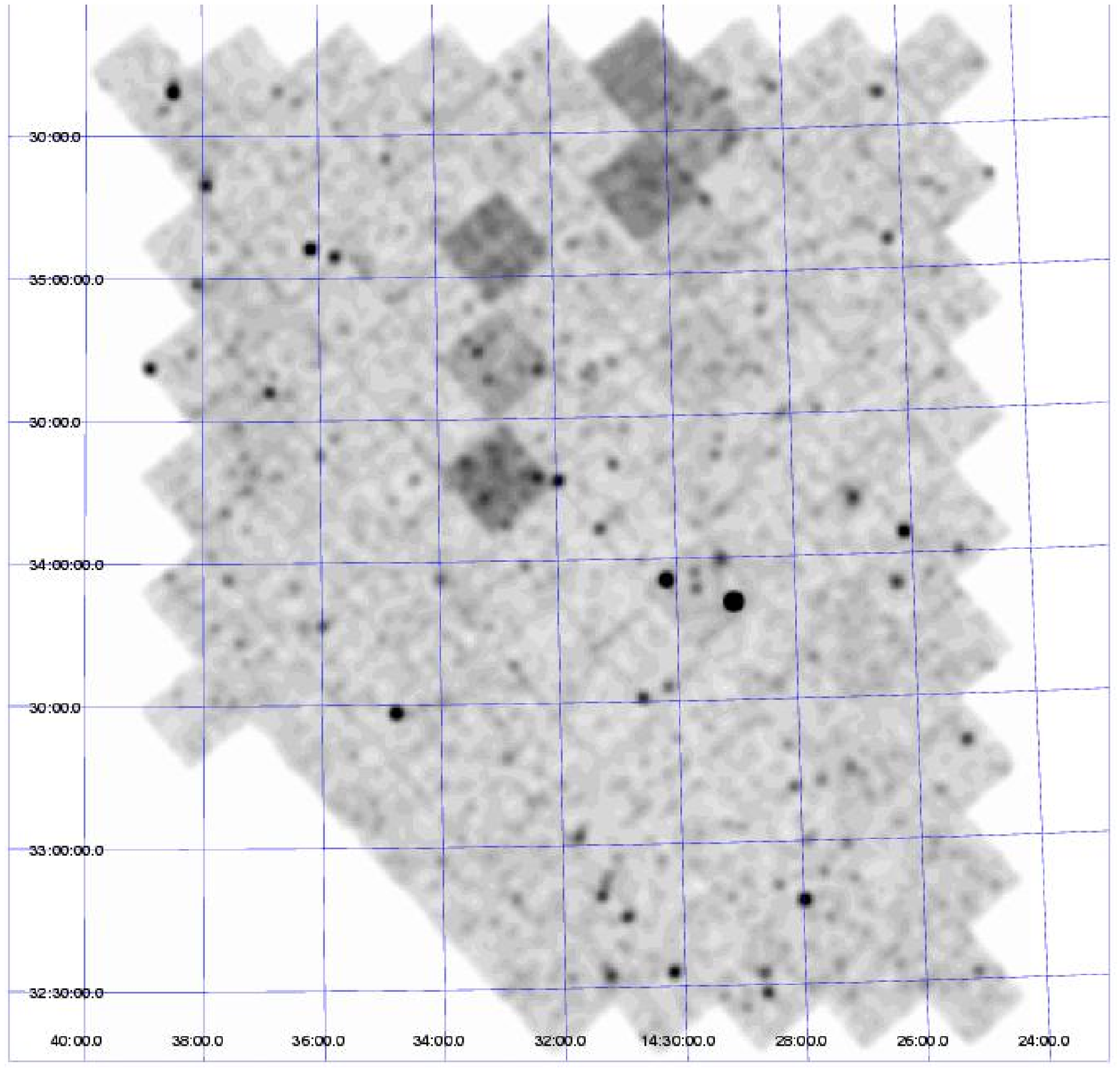}
 \caption{  A smoothed ($\approx 60\farcs$ Gaussian) coadded  image of the
XBo{\"o}tes survey region. Note the  six ACIS-I pointings with
enhanced background ($\sim 2\times$ average background levels).
 The detection of  point sources is flux-limited and unaffected by the
background even in these higher background pointings.
}
\label{fig:raw_smo_ff}
\end{figure}

For the purpose of analysis the data were
filtered into the following energy bands:
0.5--7, 0.5--2 and 2--7~keV. The energy
bands were chosen to allow comparison to ROSAT results
and to be consistent with other X-ray surveys. 
    Events with E$>7$~keV are dominated by the particle background, in
    large part because of the rapid decline in telescope effective area at these
    higher energies.  By eliminating events with E$>7$~keV, we can
    reduce the background considerably, without significantly reducing 
    the number of source photons.
The background in the 0.5--7~keV band is a factor
of $\sim 4$ lower than the full 0.5--10~keV band data, whereas the number of source photons, assuming
a typical AGN type power law spectrum, is virtually unchanged.
        Typical background counts per observation and per ACIS-I field were ~2650
in the 0.5--7~keV band, 900 in 0.5--2~keV band  and 1750 in
the 2--7~keV band.
These values correspond
to $\sim$ 2.8, 0.96, and 1.9~$\times 10^{-3}$ events  arcsec$^{-2}$.
The entire field showing the six higher background observations
is shown 
in Figure~\ref{fig:raw_smo_ff}.
Even for the six higher background rate observations,
 the number of background counts in any putative  point 
source in any of the bands is negligible;
the data are flux, not background limited.
Extended sources, such as clusters
of galaxies,  were searched for in the 0.5--2~keV band where
the lower background provides enhanced sensitivity
and where we expect most of the source counts from the
thermal bremsstrahlung emission.

\begin{figure}
\plotone{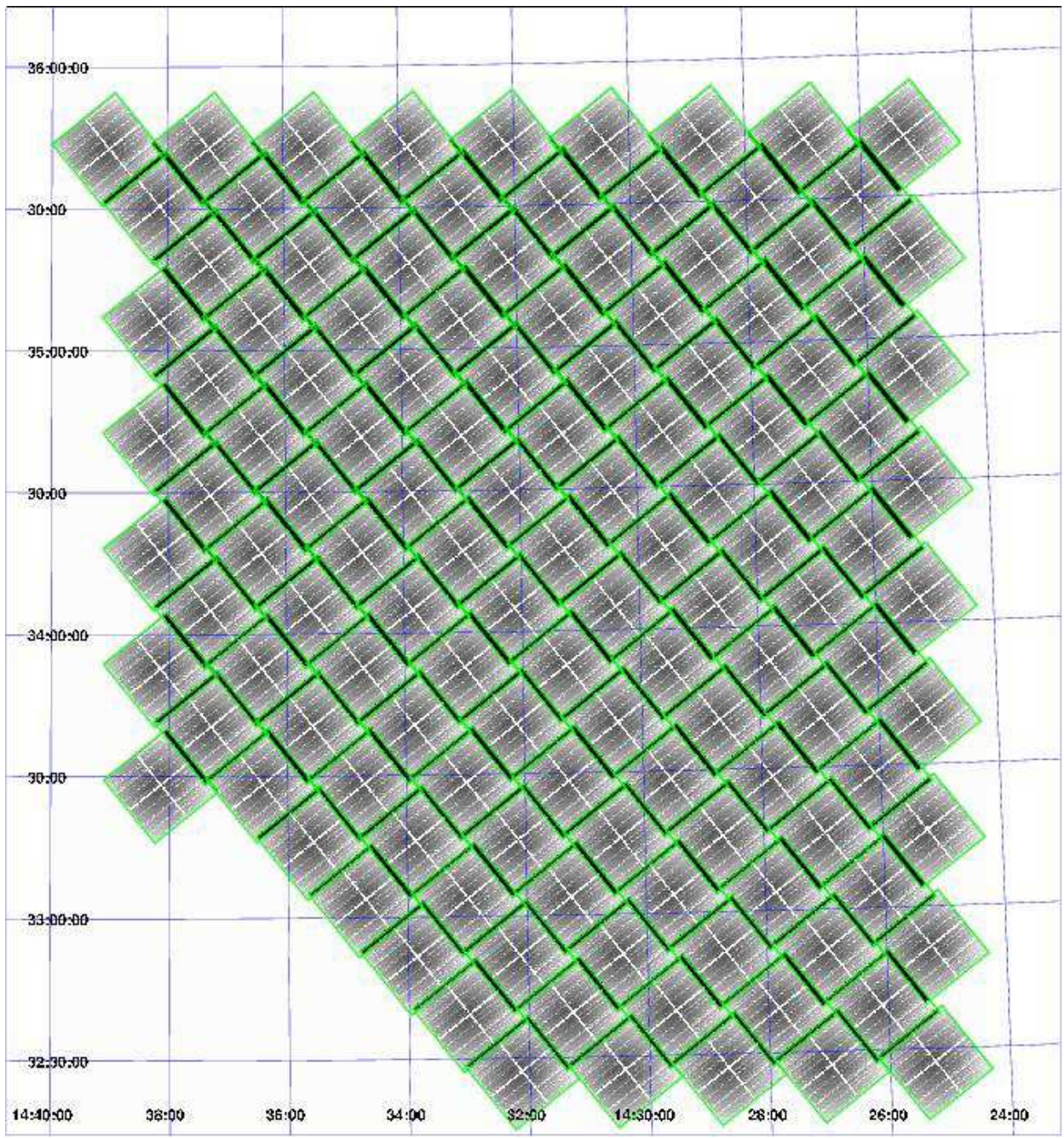}
      \caption{
The effective area map at 1.5keV for the XBo{\"o}tes survey where darker
colors indicate larger effective areas. The prominent
features are the 126 ACIS-I fields each comprised of 4 ACIS-I CCDs,
a white cross pattern of  reduced sensitivity in the gaps between the
CCD chips that are filled by dithering during the observations and
black edges where the fields overlap. The effective area peaks at
the aim point of each ACIS-I field and drops by roughly 20\% near
the edge.
  }
\label{fig:expmap}
\end{figure}

\begin{figure}[p]
\plotone{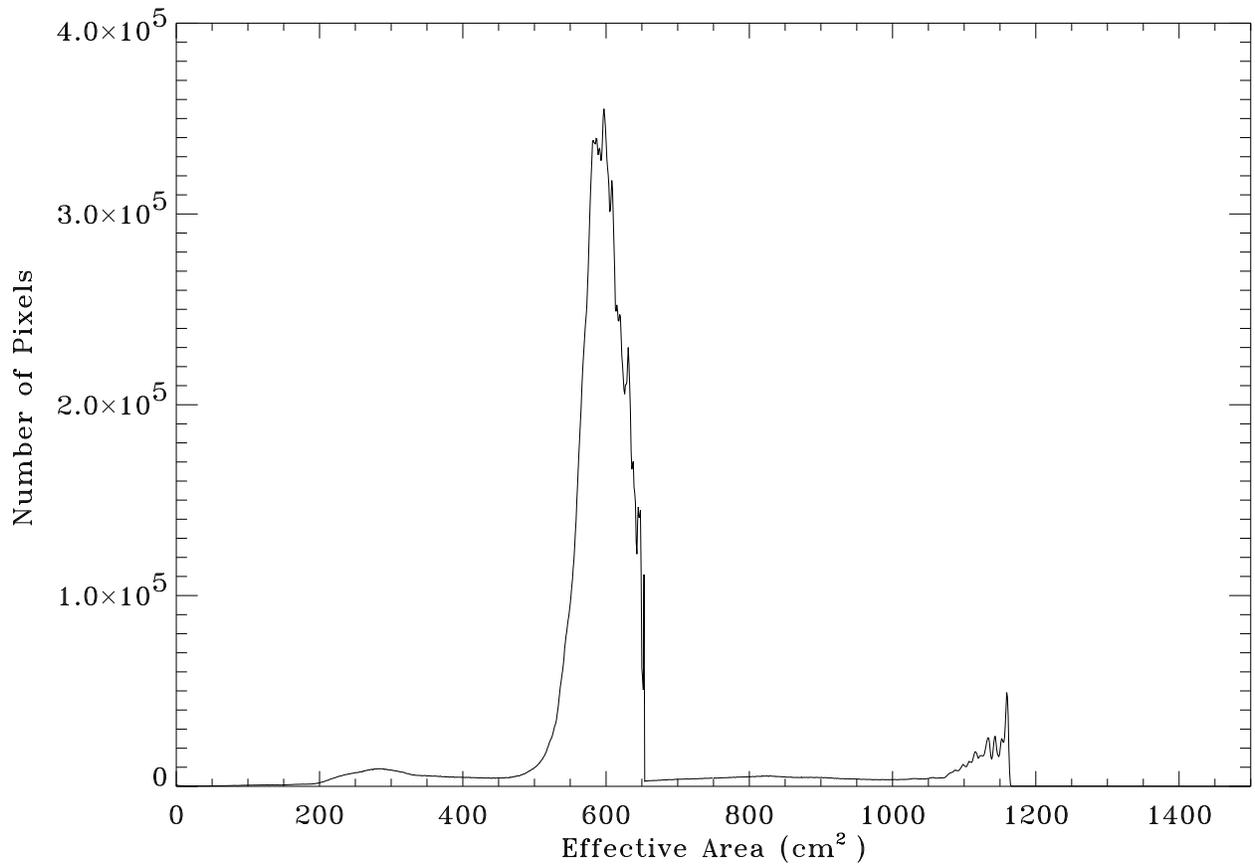}
 \caption{ 
A histogram of the effective area at 1.5~keV.  The equivalent of
13 ACIS-I fields (11\% of the area) lies in overlap regions, while
3-5\% lie in low exposure areas. The low exposure areas are created by the gaps between the
four ACIS-I CCDs and the edge of the detectors that are partially
 filled by dithering during the
exposure.}
\label{fig:EA_hist}
\end{figure}

An exposure-map, representing  the effective area at 1.5~keV of the entire field,  was generated
using the standard  \verb+ CIAO+ version 3.0 data processing and analysis tools \citep{CIAOman}.
The exposure map and its histogram are  presented in Figures~\ref{fig:expmap}
and \ref{fig:EA_hist}.
The exposure map was used to verify the overlap of the separate
observations and  to normalize the flux of detected sources.
In addition, the exposure map was used as an  input mask for 
Monte Carlo simulations. The exposure map and histogram
verify that the fields overlap each other by $\sim 1$ arc-minute.
Telescope vignetting causes the effective exposure to peak on
axis and to drop by $\sim 20 \%$ near the edge of the field. 
The exposure map histogram shows that
the overlap region corresponds to $ \sim 11 \%$ of the survey area, or
the equivalent of  $\approx 13 $ 
ACIS-I fields. This overlap is  distributed in a complicated web-like structure over
the entire survey. Although this overlap region provides a significant
area with $\approx 2$ times deeper exposure, its distribution represents
a complexity in the analysis and a potential source of systematic error in
our search for large scale structure. For this paper, we have
chosen  to perform source detection on an ACIS-I field-by-field basis.
Due to the overlap of the observations, some sources are detected
independently in separate pointings. In these cases, the detection
with the smaller off-axis angle, and hence the  smaller Point Spread Function (PSF),
is selected.
In this way the observations are combined, so as to maintain as close to uniform sensitivity
as possible. 
The effective area histogram (Figure~\ref{fig:EA_hist}) further shows that $3-5 \%$ of the survey
($\approx$ 5 ACIS-I fields)
corresponds to regions of low coverage. This low coverage area is
 due to inter-chip gaps and detector edges that have been 
smoothed by the observatory dither.  This  low exposure area  causes
a minor loss of sensitivity in the survey; a 
small fraction of the sources in our survey may be missed and
some source fluxes, when corrected for exposure, have a larger uncertainty.

\section{Source Detection and Properties}
Images of the fields were created with a spatial scale of $1\farcs968~$ per pixel
($4 \times 4 $ CCD pixels)
in the  0.5--7~keV energy band.
Sources were detected with 
the \verb+ CIAO+ version 3.0 \verb+ wavdetect+ 
 software source detection package \citep{Freeman}.
For point source detection, the wavelet scales
searched were $\{1,2,4,8\} \times$~$1\farcs96$.
  A \verb+wavdetect+ threshold of $5 \times 10^{-5}$ was chosen to
provide a more complete catalog for optical follow-up observations
while minimizing the number of spurious sources.
 This  threshold corresponds to the
probability of incorrectly identifying a pixel
as belonging to a source.
 The  source list was further restricted  by only selecting
sources with n~$\ge$~4 counts.
At this probability threshold, and with the 
 n~$\ge$~4 count selection criteria, 3293 point sources
were detected in the 0.5--7~keV band.
For a 5000 second on-axis observation and assuming a canonical unabsorbed AGN type
x-ray spectrum, a 4 count source corresponds to a
 flux  of $ \sim 4.7 \times 10^{-15}$~ergs~cm$^{-2}$~s$^{-1}$
 in the soft (0.5--2~keV) band,
$\sim 1.5 \times 10^{-14}$~ergs~cm$^{-2}$~s$^{-1}$ in the hard (2--7~keV)
band and $\sim 7.8 \times 10^{-15}$~ergs~cm$^{-2}$~s$^{-1}$ in the full (0.5--7~keV) band.
The median source in the full (0.5--7~keV)
band
 corresponds to a flux of $1.2 \times 10^{-14}$~ergs~cm$^{-2}$~s$^{-1}$.

\subsection{Spurious Sources}
\begin{figure}[p]
\plotone{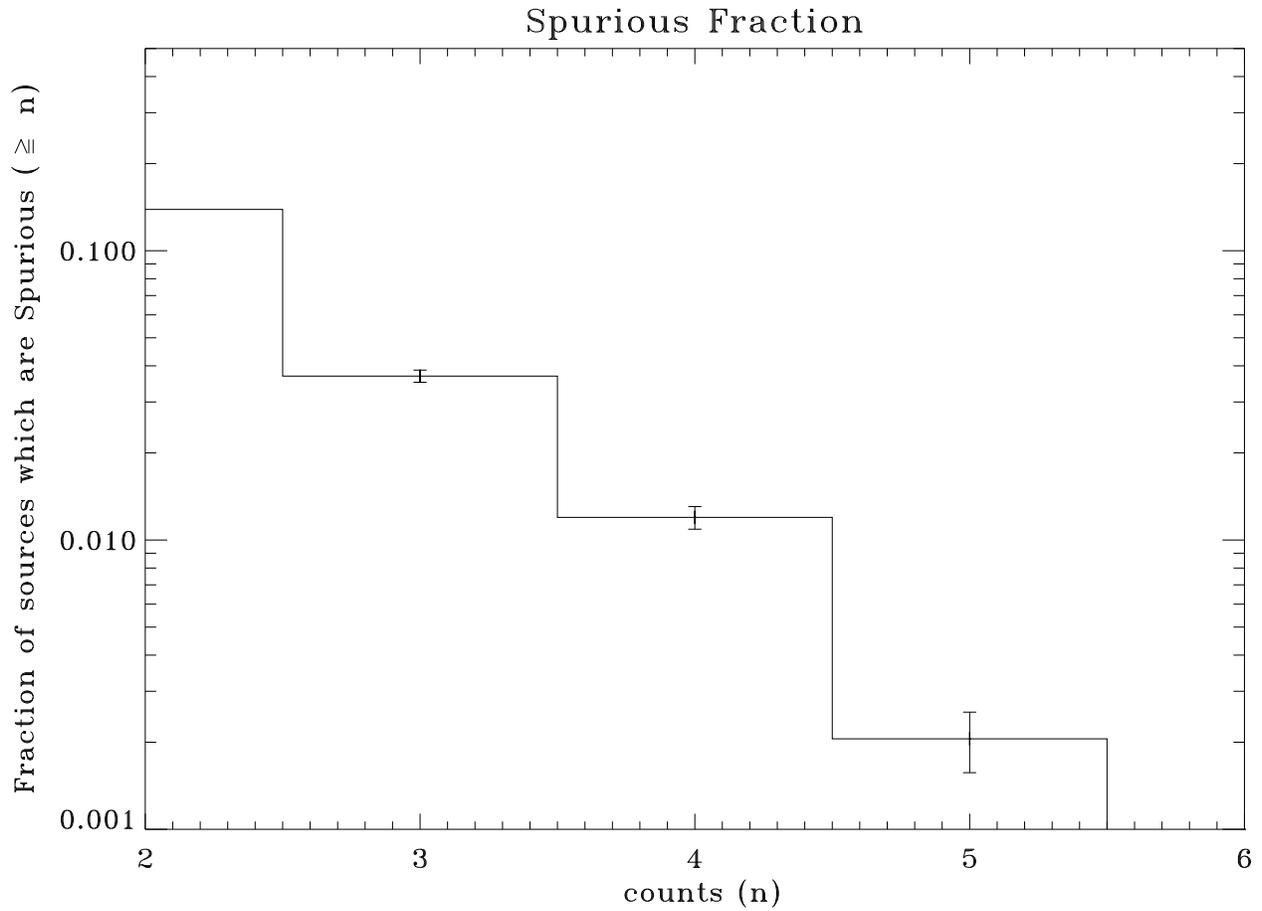}
  \caption{ Fraction of sources that are spurious as a function
of source counts. For sources $\ge$ 4 counts, the fraction is $\sim 1\%$.
Number of spurious sources was determined by the
 analysis  of  four hundred simulated   5000 second
observations generated from archival ACIS-I background data.}
\label{fig:falsies}
\end{figure}

We investigate the detection of spurious sources  
by analyzing archival ACIS-I
background data. These background ACIS-I data
were obtained with the instrument in
a similar physical state and temperature as during
the actual observations. These background data
consist of observations
of uncrowded source fields with detected sources removed, the  remaining events
are then randomized in sky coordinates, by using the  aspect solution
of the actual Bo{\"o}tes observations. 
Details describing 
the  use of archival ACIS background data can be found in \citet[and references therein]{mximbkg}.
Four hundred simulated source-free ACIS-I observations were generated, each with  
 a background  rate comparable to the ACIS-I observations in each of the 
energy bands. We then analyzed these
source free background data 
using the same techniques and thresholds that were
used to analyze the Bo{\"o}tes data.
Based on our analysis of source free ACIS-I data and the
Bo{\"o}tes data fields, we expect $\sim 1\% $ spurious or $\sim 35$ sources
with  $\ge $ 4 counts in our final  source list. 
 The fraction of spurious sources as a function
of source counts is presented in
Figure~\ref{fig:falsies}.
Similar Monte Carlo simulations performed for
a wavdetect probability threshold of $1 \times 10^{-6}$
resulted in only four spurious sources with $\ge 4$ counts.

\begin{figure}[p]
\plotone{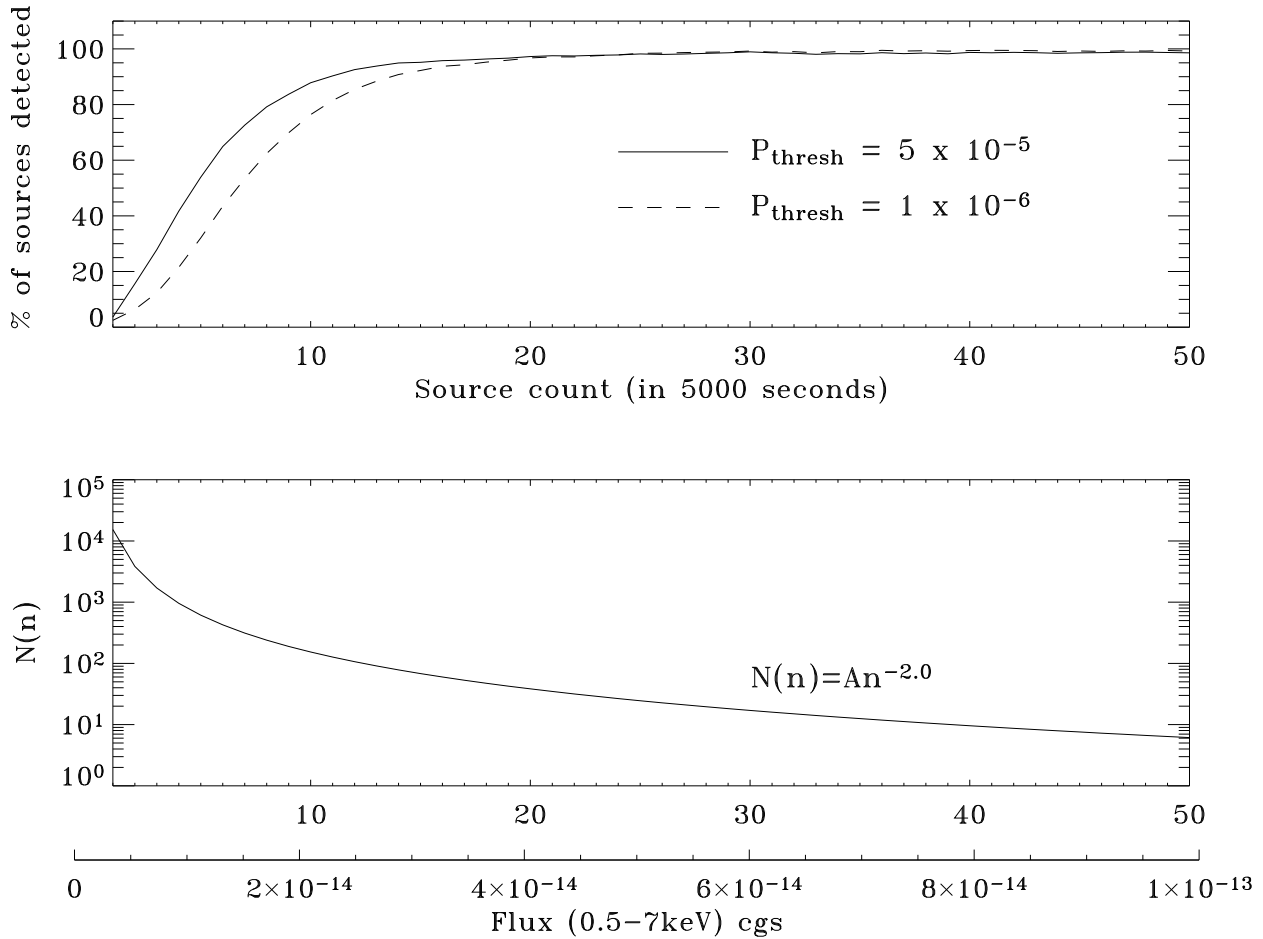}
      \caption{
 Estimates of the point-source completeness  and sensitivity
      of the XBo{\"o}tes survey based on Monte Carlo simulations.
    As part of these simulations we examined different wavdetect
    thresholds  with results shown for $P=5\times 10^{-5}$ (solid)
    and $P=10^{-6}$ (dashed curves) in top panel.  Because our observations are
    flux rather than background limited, raising the threshold to
    $5 \times 10^{-5}$ adds roughly 800 to the sample while introducing
    only about 30 spurious sources.  The simulations assumed a power-law
    distribution of sources with $N(S)= A/S^{-2}$ sources per flux
    density interval (bottom panel).  Abscissa represents
source counts for 5000 second exposure. To 
calculate flux multiply rate by ECF.}
\label{fig:completeness}
\end{figure}

Using a MARX \citep{MARXMN}
based Monte Carlo simulation program, we have examined
the fraction of sources that are detected as a function
of source flux (counts). The results of these simulations
are presented in Figure~\ref{fig:completeness}. These
simulations were carried out with appropriate values
of background and with two different wavdetect
probability thresholds, one threshold, $ 5 \times 10^{-5}$, is applicable 
to our Bo{\"o}tes source list. The other threshold of  $ 1 \times 10^{-6}$
 was analyzed  to examine the tradeoffs between 
source detection and  spurious sources.
The $ 5 \times 10^{-5}$ threshold is more sensitive to low count
sources and this  difference in sensitivity 
is  magnified
by the power law nature of the
number of sources per flux density interval. 
For these simulations we have used a appropriately normalized 
 power law ($N(n)=An^{-2.0}$), describing the
number
of sources per flux density interval to illustrate the benefits
of the $P=5\times 10^{-5}$ threshold. 
 Analysis of the
simulated data shows that 
when comparing the two different thresholds, the 
$ 5 \times 10^{-5}$ probability threshold  detects $\approx 800$ more 
sources, with the expectation that only $\approx$ 30 are spurious.

\subsection{Extended Sources}
\begin{figure}[p]
\plotone{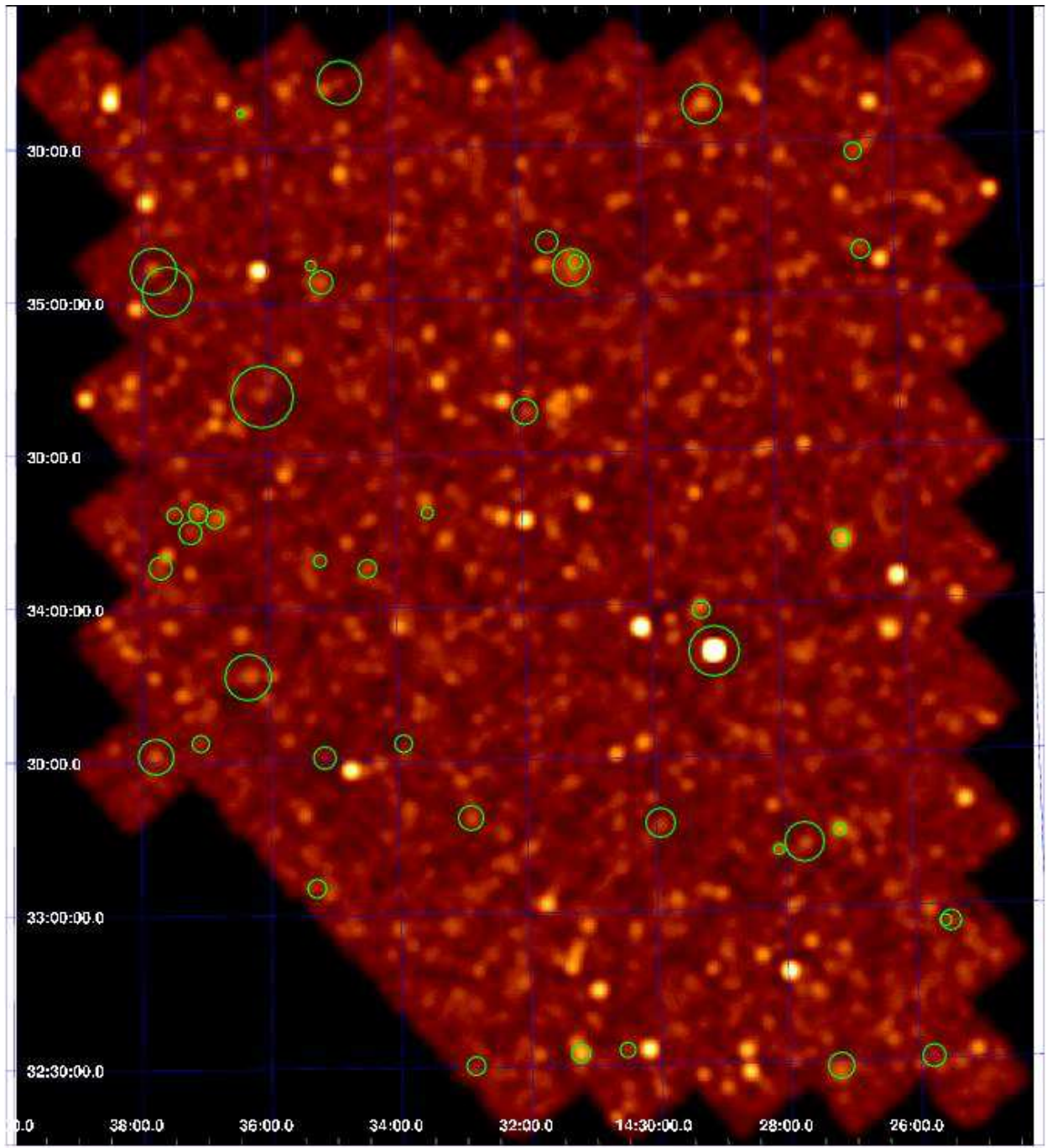}
  \caption{A smoothed ($\approx 60\farcs$ Gaussian), processed image showing locations of 43 extended
sources. The circles marking clusters are ten times the size of the detected source.
  Point sources are not marked.}
\label{fig:smo_color}
\end{figure}

Extended sources were detected
with wavelength scales
larger than the PSF using the \verb+zhtools wvdecomp+ \citep{1998ApJ...502..558V}
 software package.
In addition,
all detected sources were compared  with the local PSF. The
profile of a detected source was fit to a Gaussian with the 
width fixed at the local equivalent PSF. The fit was
then repeated with the width a free parameter. If the fit with
the Gaussian fixed width
 was consistent with the free parameter fit, the source
was deemed point-like.
The detection algorithm is similar to the 
technique described in \citet{VikCLUST}.
Forty three extended sources were detected at a
existence significance
threshold equivalent to $\approx$~3~$\sigma$.
The majority of these extended sources are expected to
be clusters of galaxies however some fraction
may be nearby star-forming galaxies.
Detecting extended sources is a complicated function
depending on the flux, extent of the source and the local size
of the PSF. We estimate that our on axis detection limit
is $\approx 1 \times 10^{-14}$~cgs (0.5--2keV) for sources that are 
just demonstrably larger than the PSF. 
  Based on  
the Log N - Log S
for clusters of galaxies reported in 
\citet{VikCLUST} and \citet{Rosati} and our flux limit, our $9.3$~degree$^{2}$ survey
is less than 50 \% complete.
We estimate that 
our effective surveyed area at fluxes  of $1\times 10^{-14}$ is only $\approx 4$~degree$^{2}$. 
The search for
extended sources was done using the 0.5--2~keV band images,
which have $ \sim 3 \times $ less background than the
full 0.5--7~keV band data. The extended source
properties are presented in Table~\ref{tab:extended} their
positions in the XBo{\"o}tes field are shown in Figure~\ref{fig:smo_color}.
The cluster sample from this survey will be explored in detail in
future publications.

\subsection{Point Source Properties}
\verb+Wavdetect+  was used  only as a source detection algorithm
because on binned images it can lead to incorrect
source positions and positional errors.
For each detected source we generated a full resolution image 
($0\farcs492$/pixel) of the region around the source and then recalculated
the source position, flux and their uncertainties. As part of conducting the optical 
matches in \citet{KateB}, we verified that this method was more reliable than
simply using the initial results from \verb+wavdetect+. 
The counts for each source  were calculated using aperture photometry by summing
 counts within the $90 \% $ Encircled Energy (EE) radius
of the centroid determined in the full resolution image around each  source position. Background
values were determined 
from an average value for each observation, with
the  fluxes of the detected sources removed.

We defined the uncertainty in the position of an $n$ count source as 
$r_{50}/\sqrt{n -1}$ if  $n\ge 5$  and $r_{50}$ if $n < 5$ where $r_{50}$  is the 50 $\%$ EE 
radius of a Gaussian approximation to the Chandra X-ray Observatory (CXO) PSF.
Source positions and positional errors 
were derived using  the  0.5--7~keV band,  where the
statistics were best. Details of the source photometry and source
locating procedure are presented in Paper I.

Estimates of the source counts and fluxes for on-axis sources  
with greater than $\gtrsim$~250 counts ( count rates $\gtrsim 5 \times 10^{-2}$ s$~^{-1}$) 
are less reliable because the ACIS pile-up phenomenon leads to the migration
of events into bad grades that are rejected.
Detections that suffer
from event pile-up  also systematically bias spectral information such as hardness ratio at even
lower count rates.
Very faint mode and de-streaking
further exacerbate the effects of event pile-up.
 The effect of pile-up is most pronounced for sources close
to ``on-axis'' and decreases rapidly with off-axis angle. 
The cause and effects of ACIS pile-up are described in detail
in the Chandra Proposer's Guide \citep{POG}.

Four counts  in  5000 seconds corresponds to  a flux of
  $ \sim 4.6 \times 10^{-15}$~ergs~cm$^{-2}$~s$^{-1}$
 in the 0.5--2~keV band,
$\sim 1.5 \times 10^{-14}$~ergs~cm$^{-2}$~s$^{-1}$ in the 2--7~keV
band and $\sim 7.2 \times 10^{-15}$~ergs~cm$^{-2}$~s$^{-1}$ in the 5--7~keV.
These flux values have been calculated assuming Galactic absorption ($NH=1\times 10^{20}$),
an on-axis point source,
and a common X-ray AGN source power law
spectrum with photon index $\Gamma = 1.7$.
The on-axis Energy Flux Density to Count Rate Conversion Factors (ECF),
presented in Table~\ref{tab:ECFs}, were obtained from the
Portable Interactive Multi-Mission Simulator 
\citep[PIMMS]{PIMMS}. The ECFs were generated for the above spectral
properties and for the March-April 2003 time period of
these observations.
The conversion of counts to flux in the various bands is
done for the benefit of comparison to other surveys; the actual
flux and flux limit for each source depends on the
 specific  source spectrum.
The cgs flux values have been calculated for 
an on-axis source  and for a common 5000 second equivalent
exposure.

\begin{figure}[p]
\plotone{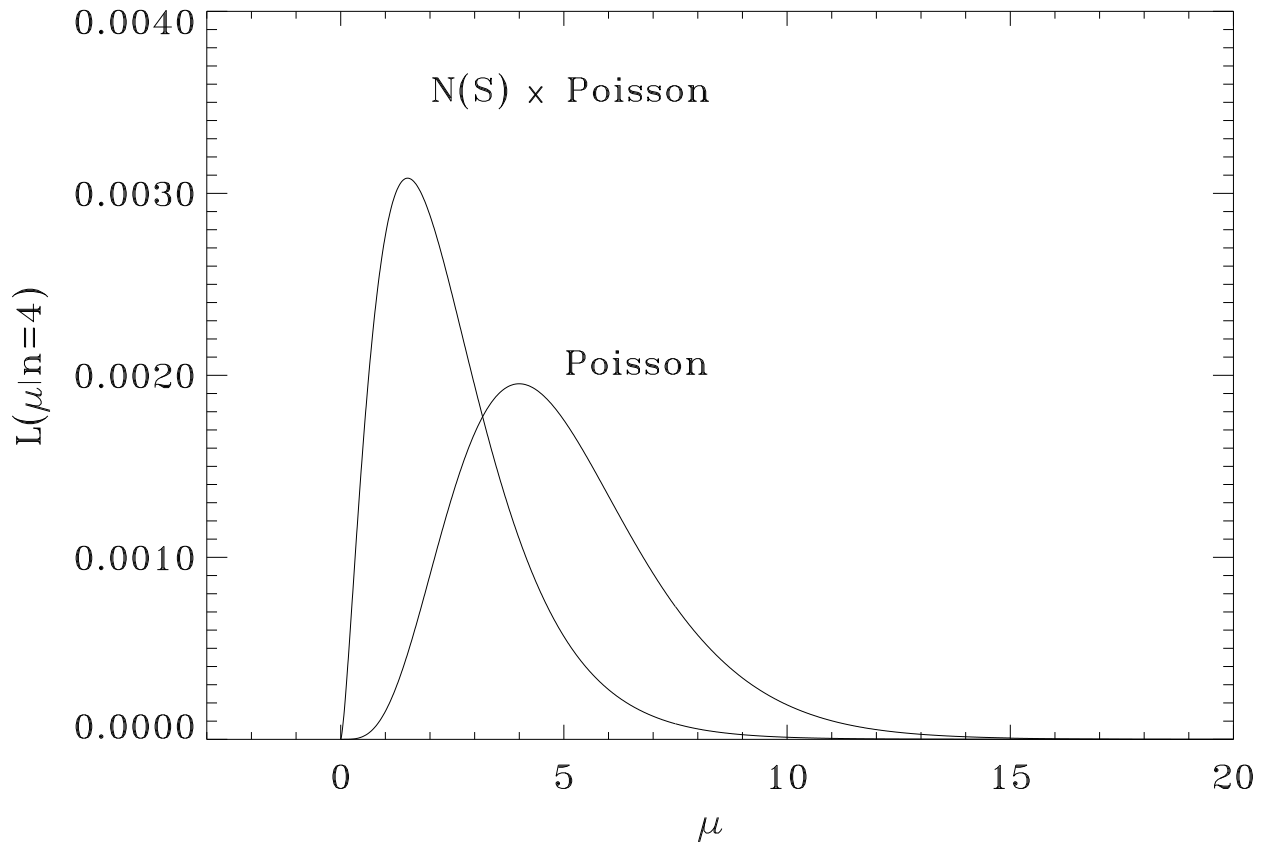}
 \caption{ Likelihood of mean ($ \mu $) having detected $n=4$ counts.
Detecting $n$ counts from a source drawn from an underlying power law
number of sources per flux density interval, $N(S) \sim S^{-\beta}$  most likely came 
from a source with true mean counts of $ \mu = n-\beta$ where $\beta$
is the power law index, rather than the purely Poisson result that 
 $ \mu = n $. For a 4 count source
and Euclidian $N(S)$, the most likely value of $ \mu = 4-2.5 = 1.5$.}
\label{fig:NSPoi}
\end{figure}

 The reported fluxes and counts of faint sources are subject to 
large uncertainties and biases.
 It can  be shown, through
maximum likelihood or  Monte Carlo simulation, that given a
number of sources per flux density interval distribution that
is a power law $N(S)\sim S^{-\beta}$, and ignoring source confusion,
the counts detected for a source  do not represent the most probable
estimator of the true mean counts of the source. 
In fact, due to Eddington bias \citep{Eddington}, an n
count source detection 
most likely  came 
from a source with a true  mean count
of  $n - \beta$.
For $N(S)\sim S^{-\beta}$,
the probability of detecting  a source with $n$ counts is proportional to 
the product of the Poisson distribution with mean $\mu$ and the differential
number of sources per flux  density interval $ N(S) \sim \mu^{-\beta}$.
\begin{equation}
            P(n|\mu,\beta) \propto  \left(\frac{\mu^{n}e^{-n}}{n!} \right) \mu^{-\beta}
\end{equation}
Having detected $n$, the most likely value of $\mu$ can be found to be:
\begin{equation}
            \frac{dP}{d\mu}=0 \Rightarrow \mu = n-\beta.
\end{equation}
 For a simple single power law Euclidian $N(S)$, ($\beta = 2.5$) a $4$ count
source detection is therefore most likely due to an upward fluctuation of a source with a true mean
count of  $ \mu=1.5^{+1.2}_{-1.4}$ (see Figure~\ref{fig:NSPoi}). 
We only note the importance of this 
bias -- the  Bo{\"o}tes source catalog presents fluxes
estimated assuming a Poisson distribution.
 Flux values and   were calculated
using 
  \begin{equation}
          f_x = A(\hbox{area},\hbox{time}) 
           \left( { \hbox{counts} - \hbox{background} \over 
                \hbox{exposure time} } \right) \times ECF
     \end{equation}
When the  number of counts in
a source are very few $(< 5)$, one cannot assume that the Poisson distribution 
from which the counts are sampled has a Gaussian shape.
 The standard deviation (i.e., the square-root of the variance) for such low-count cases
 has been derived by \citep{Gehrels} 
and the  uncertainties in counts (flux) are  
\begin{equation}
        \sigma_{count} = 1 + \sqrt{\hbox{counts} + 0.75}
      \end{equation}

\begin{figure}[p]
\plotone{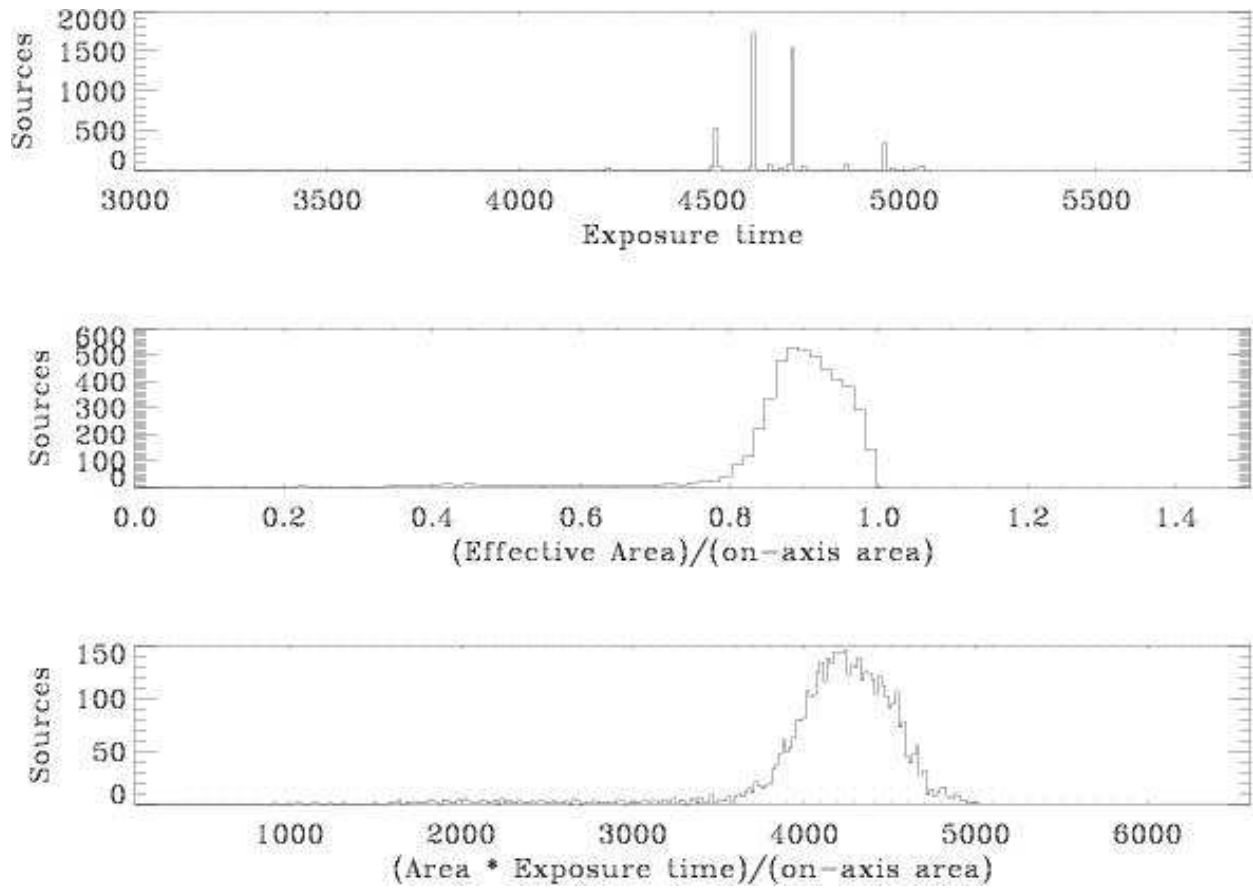}
 \caption{ 
The uniformity of the survey can be illustrated by the distribution of
   the point sources in exposure time and Effective Area (EA).  The
   exposure times varies from 4234 to 5054 seconds although the majority
   are within $\sim$10\% of each other.  The variation in the
   Effective Area (EA) for the sources
is due  primarily to telescope
vingetting and detector features that  are smoothed over
by observatory dither.
The combination of EA and exposure time variation was typically
well below $ 20 \%$. A small fraction of sources detected near edges and
chip gaps  are ``under exposed''. }
\label{fig:EA_time}
\end{figure}

    $A(\hbox{area},\hbox{time})$ corrects the observed count
    rate to the on-axis effective area and a common equivalent 5000
    second exposure time.
The  variation in exposure time and  effective area
for  all sources is typically $< 20 \%$ and is  presented in Figure~\ref{fig:EA_time}.
A small fraction of the survey near the
edge of the field of view and in the chip gaps is underexposed. 
One page of the resulting point source catalog is presented in Table~\ref{tab:tab1}.
The entire formatted source catalog is available in the on-line version of this paper.

\section{Spectral Properties; the Hardness Ratio}
Calculating spectral properties of the point sources is constrained by the shallow nature of
the survey.
The  majority of the  detected sources individually have too few counts for spectral fitting, while 
those sources with sufficient counts have rates that
cause spectral fitting to suffer from event pile-up. Spectral
distortion from event pile-up occurs  for sources with rates of $\geq 0.01$~s$^{-1}$ 
or for sources with as few as  
$\gtrsim 50$ events.
The Hardness Ratio (HR) for each source was derived to give some indication of the source's spectral properties.
We will present
spectral results of the brighter sources in future publications. 

\begin{figure}[p]
\plotone{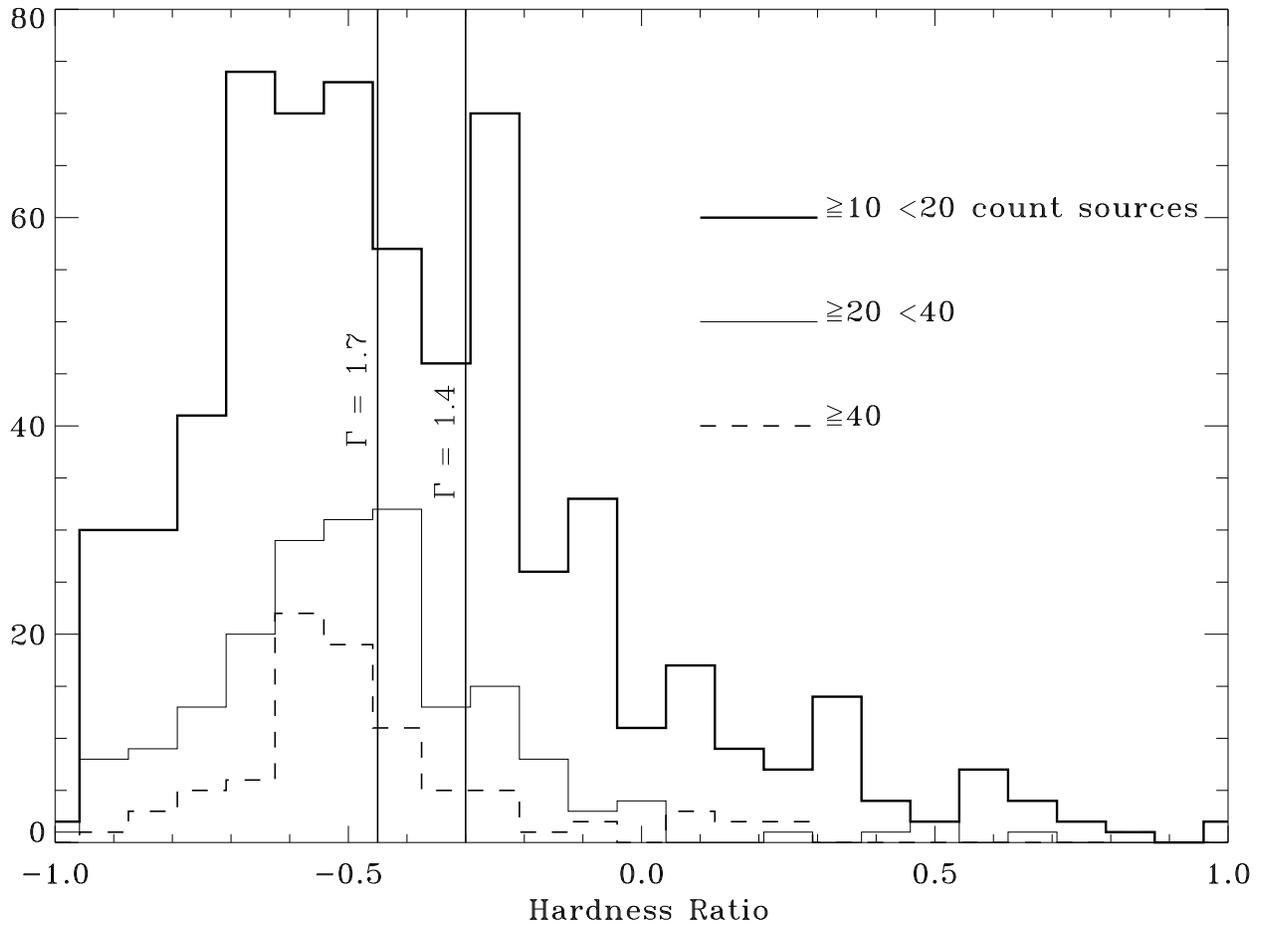}
  \caption{
     Histograms of the hardness ratios for the point sources found in the
      XBo{\"o}tes survey.  The concentration of the sources near $HR=-0.45$ is
      consistent with a source population dominated by AGN with power-law
      spectra of index $\Gamma=1.7$ combined with the expected Galactic 
      absorption (NH$=1\times 10^{20}$).  Event pile-up problems will tend to bias the HRs of
      some of the bright ($n>50$) sources upwards.}
\label{fig:hr_histo}
\end{figure}

The hardness ratio of the point sources is defined by 
$HR = (h-s)/(h+s)$,
where $s$ is the number of counts detected in the 0.5--2~keV band and $h$ is
the number of counts detected in the  2--7~keV band.
The $90 \%$ confidence limits  for the HRs were
determined using the maximum likelihood method described in the
Einstein Source  Catalog \citep{DHarris1993}. The HR and its uncertainty
for sources with $\ge$~10 events is presented in the source list. 
A histogram of the HR for sources with  $ \ge 10$ and $< 20$, $\ge 20$ and $< 40$ and $\ge 40 $ 
detected counts is presented in Figure~\ref{fig:hr_histo}.
 The $HR$ histogram indicates that, assuming a power law spectrum, some of the 
sources are soft while others are  consistent with being heavily absorbed.
 Approximately 10~$\%$ of the $\ge$~10 count sources are consistent
with  $\Gamma \sim 1.7$ and $NH >1\times 10^{22}$ (HR~$\gtrsim$~0.1)
 and $\approx$ 20~$\%$ of the $\ge$~ 10 count sources are consistent
with  $\Gamma \gtrsim 2.5$ and Galactic absorption (HR~$\lesssim$~-0.7).
 There is a clustering
of sources near $HR=-0.4$ to $-0.5$ compatible   with an 
average  power law spectrum with $\Gamma \sim 1.7$ and Galactic  absorption.

There is no conclusive evidence that the hardness ratio changes with flux
over the flux range which this survey samples. \citet{Giacconi} report
that HR increases with decreasing flux, however the effect is reported
for fluxes well below our threshold. 
We expect event pile-up to systematically bias the HR of some of 
the $\geq 40$ count sources
upward.

\section{The Number of Sources per Flux Density interval N(S)}
Calculating the number of sources per flux density interval
is subject to many systematic and statistical biases.
These biases are discussed in detail in \citet{Has93},\citet{Vik95}, 
\citet{Murd73} and \citet{SnM86}. 
Due to the uniform coverage of this survey, we use the
technique described in \citet{2003ApJ...584.1016K}. This technique
inherently accounts for statistical and instrumental biases
and it  
calibrates the photometry of the  source
detection algorithm. It allows us to use the large number of low count sources,
where the effects of the  biases are most pronounced, 
to determine the differential number of sources per flux density interval.
The differential number of sources per flux density interval
 for the soft band (0.5-2~keV) is statistically inconsistent with 
a single power-law. It is best fit by a broken power law of the
form:

\[N(S_{14})] = \left\{ \begin{array}{ll}
                      K_{1}S_{14}^{-\beta_{1}} & \mbox{if $S<S_{b}$}\\
                     K_{2}S_{14}^{-\beta_{2}} & \mbox{if $S>S_{b}$}
                     \end{array}
                  \right. \]

%
 where $S_{14}$
is the source flux in units of $10^{-14}$~ergs~cm$^{-2}$~s$^{-1}$.
 At fluxes below the break, the index is $ \beta_{1}=1.74^{+0.28}_{-0.22}$ and for brighter fluxes
it is $\beta_{2}=2.60^{+0.11}_{-0.12}$ where  $K_{1}=129^{+62 }_{-40 }$
and $K_{2} = K_{1}S_{b}^{\beta_{1} -\beta_{2}}$ is constrained by continuity at the break.
 The best fit for the break is  at $S_{b}=8.2\times 10^{-15}$~cgs,
however it is very poorly constrained
 (upper $1 \sigma$ limit $=3.0\times 10^{-14} $ and   lower $1 \sigma$ limit$ = 3.2\times 10^{-15} $).
These uncertainties  indicate statistical  $68 \%$  
($1 \sigma $)
confidence limits.
The values of the  indices and break are  in  agreement with the ROSAT  Lockman Hole  results of \citet{Has98a}
given the statistical uncertainties. The ROSAT results covered a similar
flux range and are the most directly comparable. The XBo{\"o}tes soft (0.5--2keV) band
low flux index is consistent with deep Chandra  surveys although
the flux overlap is minimal. \citet{Brandt} found integral index
of $\alpha=0.67 \pm 0.14$ for the Chandra Deep field North (CDF-N).
 Similarly, in the CHAMP survey \citet{DWKim} found integral index of  $\alpha=0.7 \pm 0.15$.

\begin{figure}[p]
\plotone{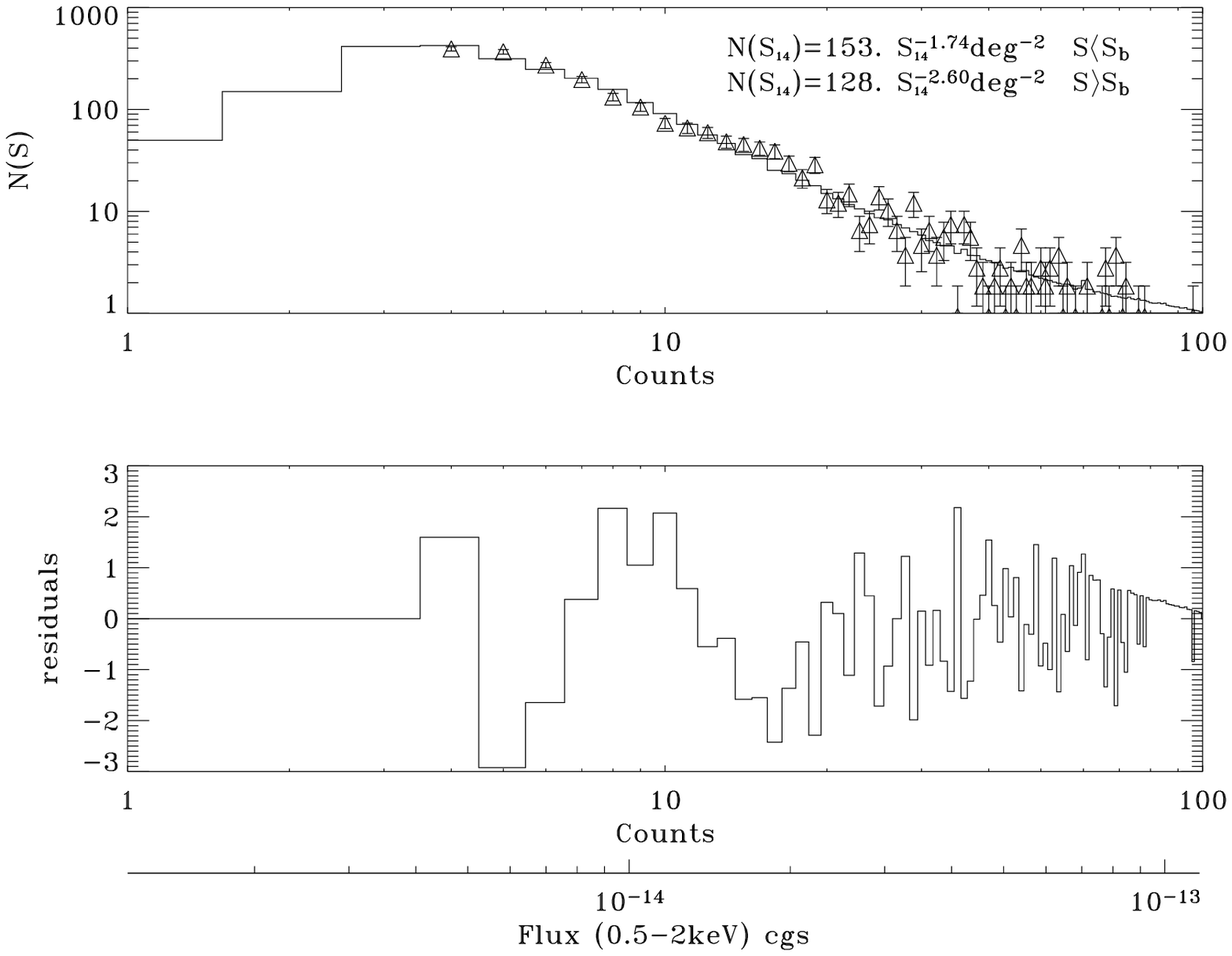}
  \caption{
 The number of sources per flux density interval in the 0.5--2.0~keV
    band.  The points show the measured counts and the histogram shows
    the best fit model. Best fit model is calculated  using 
technique described in \citet{2003ApJ...584.1016K}. 
}
\label{fig:Nn_dat_0.5-2keV_plot}
\end{figure}

\begin{figure}[p]
\plotone{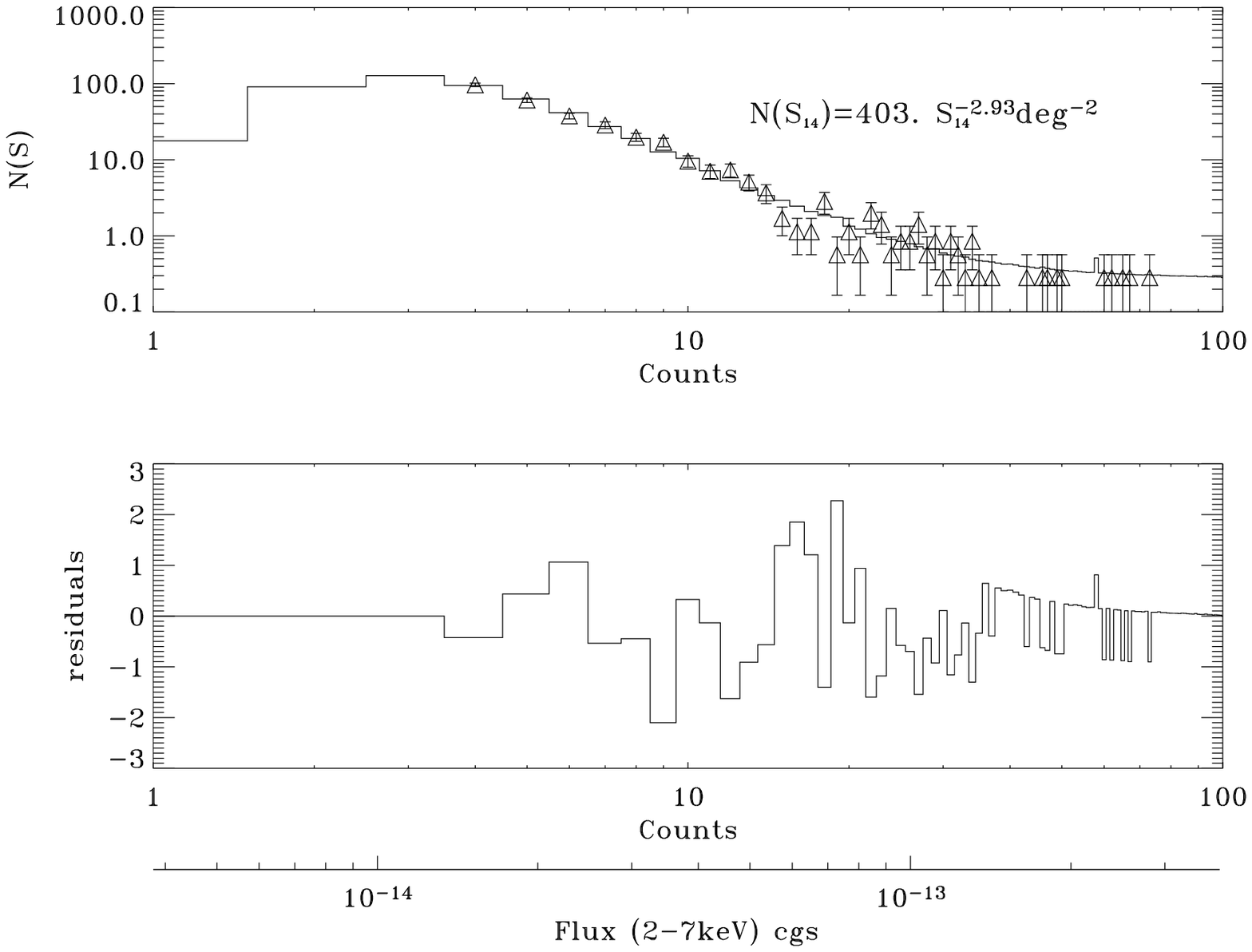} 
  \caption{
 The number of sources per flux density interval in the 2.0--7.0~keV
    band.  The points show the measured counts and the histogram shows
    the best fit model. Best fit model is calculated using 
technique described in \citet{2003ApJ...584.1016K}. 
}
 \label{fig:Nn_dat_2-7keV_plot}
\end{figure}

\begin{figure}[p]
\plotone{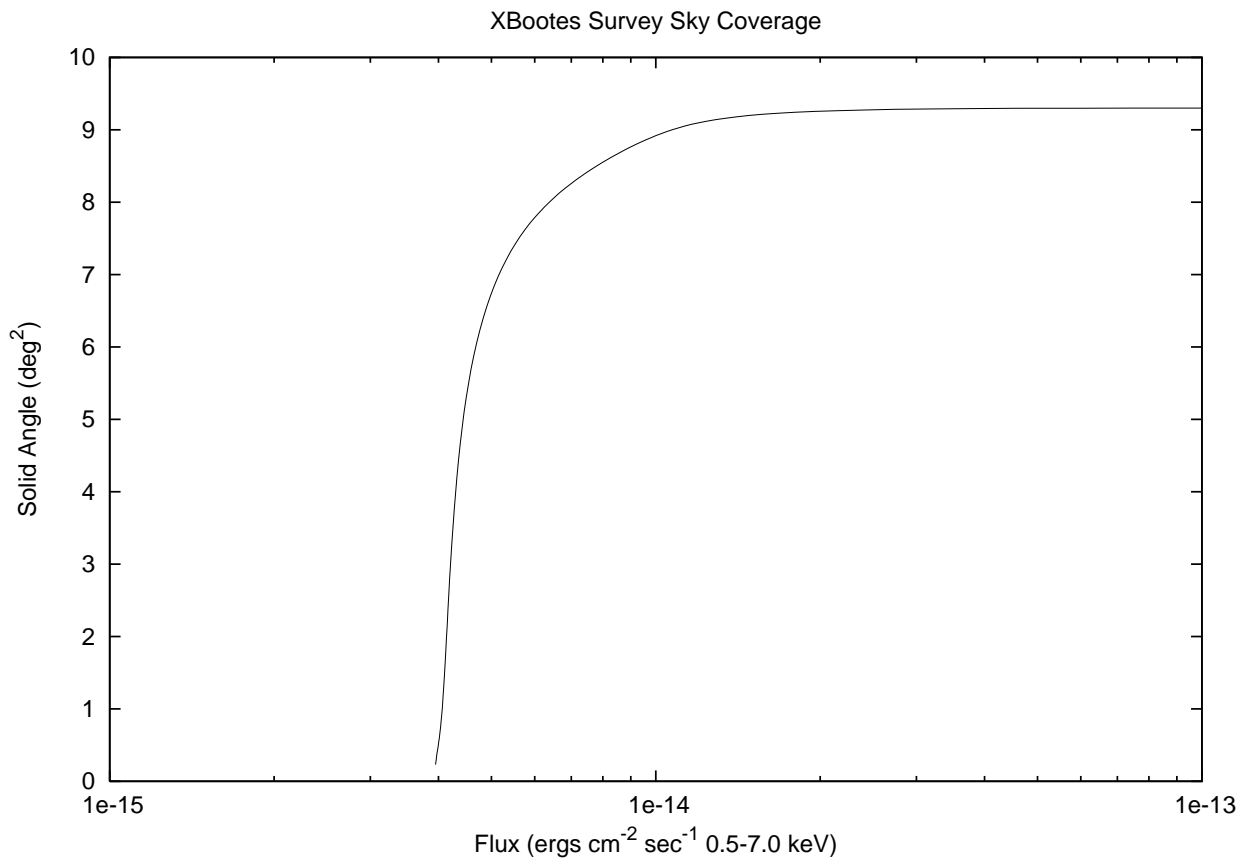}
  \caption{
Plot of the sky coverage for the XBo\"otes survey. The plot
gives the solid angle (in square degrees) as a function of the Total
band (0.5-7.0 keV) flux in $\mathrm{erg\, cm^{-2}sec^{-1}}$. The
nearly uniform exposure times of the 126 ACIS fields, accounts for
the rapid rise in coverage at the lowest fluxes. The roll over between
about 7 and 9 square degrees is due to the effects of vignetting and
the growth of the Chandra point spread function (PSF) at large off-axis
angles.}
 \label{fig:eff_sky}
\end{figure}

The differential number of sources per flux density interval
 for the hard band (2--7~keV) is  well approximated by a single power-law of the form
  $N(S)=  KS_{14}^{-\beta}$.
\begin{equation}
N(S_{14})=  [403^{+52}_{-45}] S_{14}^{-2.93(+0.09/-0.09)}\hbox{deg.}^{-2} 
\end{equation}
This slope is steeper than  the $2.65 \pm 0.1 $  value for 2--10~keV sources reported
 by \citet{Giommi}. 
These calculations assume a single average power law energy spectrum for  all the sources.
The true 
conversion of  detected counts to flux varies from source to source and this
variation would contribute to our uncertainties.
The results of our technique to determine
the number of sources per flux density interval  
are presented in Figures~\ref{fig:Nn_dat_0.5-2keV_plot} and \ref{fig:Nn_dat_2-7keV_plot}.
The effective sky coverage for point sources in this XBo{\"o}tes survey
is presented in Figure \ref{fig:eff_sky}.

\section{Summary and Conclusions}

We have analyzed the largest contiguous Chandra field with
uniform coverage and arcsecond resolution.
These results and statistics
 are dominated by sources with fluxes $\sim 10^{-14}$~cgs.
This survey is  several orders of magnitude shallower and several
orders of magnitude wider than the deepest surveys.
We have detected 4642 sources with $\ge 2$ counts. Our $99 \%$ reliable
source list consists of  $ 3293 $ point sources in the full (0.5--7~keV) band.
The resulting source list is ideal for multi-wavelength follow-up observations.

The differential 0.5--2~keV  number of sources per flux density interval
distribution, $N(S)$, is best fit by a broken 
power law. The parameters of this fit 
are consistent with  reported ROSAT results that  covered approximately
the same $\sim 10^{-14}$~cgs 
flux regime. 
The differential 2--7~keV  $N(S)$ distribution is consistent with
a single power law with an index that is steeper than Euclidian.

In addition to point sources we detect forty-three  extended sources, consistent with
the depth of the observations and the number counts of clusters. 
Hardness ratio distributions of the point sources are consistent with
Galactic absorption and typical AGN power law spectra.   Some sources with particularly
large HR values are candidate type II QSOs and will be
investigated further.

\section{Acknowledgements}

Funding for this research was provided in part by NASA contracts
NAS8-38248, NAS8-01130, NAS8-39073 and GO grant GO3-4176A.
Funding for this research was also provided by
the National Optical Astronomy Observatory,
which is operated by AURA, Inc., under a cooperative
agreement with the National Science Foundation.
We would like to thank the anonymous referee for helpful suggestions
that improved this paper.

\pagestyle{empty}
\footnotesize
\begin{center}
\begin{landscape}
\begin{longtable} {|c c c r r r r r | }\hline
CXO Name & RA$_{2000}$ & DEC$_{2000}$ & Positionl Error&  Size & Size Error& Net Counts\hphantom{} & $S_{14}$ Flux \hphantom{0}   \\ 
 &  &     &  ('')\hphantom{00000} & ('') & ('')\hphantom{00}& (0.5-2keV) &  (0.5-2keV)   \\ \hline \hline 
CXOXB J142527.8+325646&14 25 27.86 & 32 56 46.46 & 4.04\hphantom{0000} & 12&4.5\hphantom{00}&8.80$\pm$4.1&1.06$\pm$0.50   \\
CXOXB J142532.9+325644&14 25 32.90 & 32 56 44.75 & 2.12\hphantom{0000} & 6.9&3.2\hphantom{00}&10.5$\pm$4.3&1.27$\pm$0.53   \\
CXOXB J142547.5+323025&14 25 47.51 & 32 30 25.30 & 3.16\hphantom{0000} & 13&4.5\hphantom{00}&17.9$\pm$5.3&2.02$\pm$0.60   \\
CXOXB J142632.5+350821&14 26 32.51 & 35 08 21.18 & 2.32\hphantom{0000} & 11&3.0\hphantom{00}&22.3$\pm$5.8&2.70$\pm$0.70   \\
CXOXB J142637.0+352734&14 26 37.04 & 35 27 34.57 & 3.31\hphantom{0000} & 10&4.2\hphantom{00}&9.30$\pm$4.1&1.12$\pm$0.50   \\
CXOXB J142657.9+341201&14 26 57.90 & 34 12 01.40 & 3.32\hphantom{0000} & 44&4.4\hphantom{00}&182.$\pm$14.7&22.1$\pm$1.78   \\
CXOXB J142709.3+331510&14 27 09.30 & 33 15 10.19 & 2.17\hphantom{0000} & 16&2.2\hphantom{00}&54.9$\pm$8.4&6.33$\pm$0.98   \\
CXOXB J142713.7+322857&14 27 13.78 & 32 28 57.82 & 2.96\hphantom{0000} & 15&3.0\hphantom{00}&27.0$\pm$6.3&3.11$\pm$0.72   \\
CXOXB J142741.8+331252&14 27 41.89 & 33 12 52.75 & 4.08\hphantom{0000} & 23&6.4\hphantom{00}&32.3$\pm$6.8&4.00$\pm$0.84   \\
CXOXB J142805.8+331130&14 28 05.84 & 33 11 30.88 & 1.18\hphantom{0000} & 6.3&4.3\hphantom{00}&28.1$\pm$6.3&3.48$\pm$0.79   \\
CXOXB J142900.6+353734&14 29 00.60 & 35 37 34.30 & 3.16\hphantom{0000} & 23&3.9\hphantom{00}&54.6$\pm$8.5&6.91$\pm$1.07   \\
CXOXB J142901.9+335033&14 29 01.91 & 33 50 33.89 & 4.21\hphantom{0000} & 29&5.0\hphantom{00}&49.2$\pm$8.2&6.23$\pm$1.03   \\
CXOXB J142916.1+335929&14 29 16.15 & 33 59 29.75 & 3.21\hphantom{0000} & 28&4.7\hphantom{00}&76.5$\pm$9.8&9.69$\pm$1.25   \\
CXOXB J142955.8+331711&14 29 55.87 & 33 17 11.89 & 3.66\hphantom{0000} & 17&4.3\hphantom{00}&22.0$\pm$5.8&2.72$\pm$0.72   \\
CXOXB J143031.0+323257&14 30 31.00 & 32 32 57.25 & 3.20\hphantom{0000} & 9.0&3.8\hphantom{00}&7.90$\pm$3.9&0.91$\pm$0.45   \\
CXOXB J143104.8+350714&14 31 04.83 & 35 07 14.45 & 2.13\hphantom{0000} & 8.5&2.2\hphantom{00}&15.9$\pm$5.0&1.97$\pm$0.63   \\
CXOXB J143109.1+350609&14 31 09.17 & 35 06 09.06 & 3.56\hphantom{0000} & 21&3.6\hphantom{00}&37.7$\pm$7.2&4.67$\pm$0.90   \\
CXOXB J143113.8+323225&14 31 13.81 & 32 32 25.43 & 2.57\hphantom{0000} & 25&3.7\hphantom{00}&97.5$\pm$10.9&11.2$\pm$1.26   \\
CXOXB J143131.8+351115&14 31 31.87 & 35 11 15.63 & 4.43\hphantom{0000} & 13&4.9\hphantom{00}&8.60$\pm$4.1&1.06$\pm$0.50   \\
CXOXB J143156.1+343806&14 31 56.12 & 34 38 06.65 & 2.56\hphantom{0000} & 15&5.5\hphantom{00}&34.1$\pm$6.9&4.23$\pm$0.86   \\
CXOXB J143251.5+323018&14 32 51.50 & 32 30 18.29 & 3.34\hphantom{0000} & 10&5.8\hphantom{00}&10.6$\pm$4.4&1.28$\pm$0.53   \\
CXOXB J143253.1+331806&14 32 53.14 & 33 18 06.53 & 3.50\hphantom{0000} & 15&6.3\hphantom{00}&19.1$\pm$5.4&2.36$\pm$0.68   \\
CXOXB J143330.1+341835&14 33 30.11 & 34 18 35.72 & 2.07\hphantom{0000} & 7.0&3.2\hphantom{00}&11.4$\pm$4.5&1.38$\pm$0.54   \\
CXOXB J143355.2+333328&14 33 55.20 & 33 33 28.77 & 3.41\hphantom{0000} & 10&4.4\hphantom{00}&9.10$\pm$4.1&1.12$\pm$0.51   \\
CXOXB J143427.4+340746&14 34 27.43 & 34 07 46.88 & 2.99\hphantom{0000} & 10&4.7\hphantom{00}&12.8$\pm$4.7&1.58$\pm$0.58   \\
CXOXB J143449.0+354301&14 34 49.05 & 35 43 01.77 & 4.39\hphantom{0000} & 25&6.6\hphantom{00}&34.7$\pm$7.0&4.20$\pm$0.85   \\
CXOXB J143508.8+350349&14 35 08.85 & 35 03 49.27 & 3.98\hphantom{0000} & 20&4.4\hphantom{00}&26.7$\pm$6.3&3.23$\pm$0.76   \\
CXOXB J143509.0+333050&14 35 09.03 & 33 30 50.55 & 3.28\hphantom{0000} & 13&3.9\hphantom{00}&15.9$\pm$5.1&1.93$\pm$0.62   \\
CXOXB J143511.9+340922&14 35 11.94 & 34 09 22.51 & 2.49\hphantom{0000} & 7.6&3.5\hphantom{00}&9.30$\pm$4.1&1.12$\pm$0.50   \\
CXOXB J143517.5+330518&14 35 17.56 & 33 05 18.30 & 3.95\hphantom{0000} & 10&5.1\hphantom{00}&7.60$\pm$3.9&0.94$\pm$0.48   \\
CXOXB J143518.3+350710&14 35 18.31 & 35 07 10.03 & 1.31\hphantom{0000} & 6.1&3.4\hphantom{00}&21.6$\pm$5.7&2.62$\pm$0.69   \\
CXOXB J143601.9+344226&14 36 01.94 & 34 42 26.03 & 3.66\hphantom{0000} & 17&3.9\hphantom{00}&22.3$\pm$5.8&2.70$\pm$0.71   \\
CXOXB J143615.4+334650&14 36 15.44 & 33 46 50.48 & 3.47\hphantom{0000} & 21&4.1\hphantom{00}&37.5$\pm$7.2&4.54$\pm$0.87   \\
CXOXB J143624.3+353708&14 36 24.32 & 35 37 08.85 & 0.83\hphantom{0000} & 4.3&2.8\hphantom{00}&26.6$\pm$6.2&3.22$\pm$0.75   \\
CXOXB J143651.0+341737&14 36 51.06 & 34 17 37.71 & 2.77\hphantom{0000} & 10&3.2\hphantom{00}&14.6$\pm$4.9&1.77$\pm$0.60   \\
CXOXB J143705.5+333344&14 37 05.56 & 33 33 44.67 & 2.82\hphantom{0000} & 10&4.0\hphantom{00}&12.5$\pm$4.6&1.53$\pm$0.57   \\
CXOXB J143707.0+341848&14 37 07.06 & 34 18 48.87 & 2.60\hphantom{0000} & 11&5.1\hphantom{00}&19.2$\pm$5.4&2.32$\pm$0.66   \\
CXOXB J143714.3+341503&14 37 14.35 & 34 15 03.19 & 3.05\hphantom{0000} & 13&6.8\hphantom{00}&19.5$\pm$5.5&2.25$\pm$0.63   \\
CXOXB J143729.1+341822&14 37 29.18 & 34 18 22.74 & 2.43\hphantom{0000} & 9.1&4.6\hphantom{00}&14.0$\pm$4.8&1.73$\pm$0.60   \\
CXOXB J143735.8+350214&14 37 35.87 & 35 02 14.97 & 4.97\hphantom{0000} & 29&6.0\hphantom{00}&34.5$\pm$7.0&4.28$\pm$0.87   \\
CXOXB J143742.7+340807&14 37 42.77 & 34 08 07.72 & 3.28\hphantom{0000} & 13&3.9\hphantom{00}&16.6$\pm$5.2&1.91$\pm$0.60   \\
CXOXB J143747.6+333110&14 37 47.63 & 33 31 10.41 & 3.60\hphantom{0000} & 20&3.9\hphantom{00}&33.2$\pm$6.8&4.11$\pm$0.85   \\
CXOXB J143748.4+350617&14 37 48.49 & 35 06 17.20 & 4.91\hphantom{0000} & 27&5.0\hphantom{00}&30.9$\pm$6.7&3.83$\pm$0.83   \\
\hline \hline
\caption{Properties of Extended Sources in the XBo{\"o}tes Survey.
X-ray source properties for
the forty three extended sources  detected in the Bo{\"o}tes 9.3 square
degree field. Detection was done in the 0.5--2~keV band. Source
size is from Gaussian fit to source profile.}
\label{tab:extended}
\end{longtable}
\end{landscape}
\end{center}
\normalsize
\begin{table}[p]
\begin{center}
\begin{tabular} {lccc}\hline
$\Gamma $ &ECF (0.5-2keV)  &  ECF (2-7keV)& ECF (0.5-7keV)   \\ \hline
1.4  & $5.64 \times 10^{-12}$   & $2.02 \times 10^{-11}$ & $1.09 \times 10^{-11}$\\ \hline
{\bf 1.7 } & {\bf $5.84 \times 10^{-12}$ } & {\bf$1.93 \times 10^{-11}$} & {\bf $9.81 \times 10^{-12}$  }\\ \hline
2.0  &$6.07 \times 10^{-12}$  & $1.85 \times 10^{-11}$ &$8.98 \times 10^{-12}$ \\ \hline \hline
&  &  & \\
Temp  &ECF (0.5-2keV)  &  ECF (2-7keV)& ECF (0.5-7keV)   \\ \hline \hline
{\bf 6keV } & {\bf $5.72 \times 10^{-12}$ }  & {\bf $---$}  & {\bf $---$}\\ \hline
\end{tabular}
\caption{ Conversion factors from count~s$^{-1}$ to cgs units
(ergs~cm$^{-2}$~s$^{-1}$)
for power law spectra (point sources) and thermal Bremstrahlung (extended sources).
The values are obtained from PIMMS
and are appropriate for the March -April 2003 time period of the observations.
Changing the power law index from the nominal $\Gamma=1.7$ has small effect for
soft energies ($\sim 4 \%$) and a slightly larger effect for the hard band. }
\label{tab:ECFs}
\end{center}
\end{table}

\tiny
 \begin{landscape}
\begin{longtable} {|l c c c c c  c c c c c c c | }\hline
        CXO Name &RA$_{2000}$ & DEC$_{2000}$ & $\delta$ position & counts & counts  & counts &flux($S_{14}$) & flux($S_{14}$)  & flux($S_{
14}$) & exp time & frac area & HR \\
  &  &   &  arcsec & 0.5-7keV & 0.5-2keV & 2-7keV & 0.5-7keV & 0.5-2keV & 2-7keV & & & \\ \hline \hline
CXOXB J142420.5+333922 &  14 24 20.58 & 33 39 22.13 & 5.31 & 4 & 1&3&0.88$\pm$0.70 & 0.12$\pm$0.30 & 1.39$\pm$1.29&4714.68&0.780&  \\
CXOXB J142428.1+351922 &  14 24 28.15 & 35 19 22.49 & 0.63 & 99 & 74&25&26.17$\pm$2.29 & 11.68$\pm$1.20 & 13.32$\pm$2.58&4711.62&0.831&-0.
50$^{+0.01}_{-0.01}$ \\
CXOXB J142429.6+342721 &  14 24 29.68 & 34 27 21.84 & 3.72 & 5 & 3&2&1.20$\pm$0.72 & 0.45$\pm$0.37 & 0.91$\pm$1.15&4714.68&0.837&  \\
CXOXB J142430.0+325034 &  14 24 30.03 & 32 50 34.13 & 3.73 & 7 & 6&1&1.77$\pm$0.80 & 0.94$\pm$0.45 & 0.39$\pm$1.02&4708.48&0.821&  \\
CXOXB J142433.0+342819 &  14 24 33.01 & 34 28 19.33 & 1.37 & 19 & 12&7&5.03$\pm$1.14 & 1.90$\pm$0.57 & 3.71$\pm$1.61&4714.68&0.821&-0.27$^
{+0.07}_{-0.07}$ \\
CXOXB J142434.0+331557 &  14 24 34.04 & 33 15 57.34 & 3.11 & 7 & 6&1&1.74$\pm$0.80 & 0.92$\pm$0.45 & 0.41$\pm$1.01&4714.68&0.842&  \\
CXOXB J142434.4+331340 &  14 24 34.46 & 33 13 40.88 & 0.98 & 26 & 18&8&6.59$\pm$1.29 & 2.73$\pm$0.66 & 4.06$\pm$1.69&4714.68&0.859&-0.40$^
{+0.05}_{-0.05}$ \\
CXOXB J142435.4+353855 &  14 24 35.40 & 35 38 55.00 & 2.56 & 5 & 5&0&1.04$\pm$0.67 & 0.65$\pm$0.40 & $\le$0.6&5038.95&0.859&  \\
CXOXB J142435.7+345012 &  14 24 35.75 & 34 50 12.98 & 3.12 & 8 & 2&6&1.98$\pm$0.83 & 0.29$\pm$0.33 & 3.04$\pm$1.54&4714.68&0.854&  \\
CXOXB J142435.7+354224 &  14 24 35.76 & 35 42 24.10 & 0.87 & 33 & 22&11&7.23$\pm$1.33 & 2.88$\pm$0.67 & 4.85$\pm$1.76&5038.95&0.873&-0.34$
^{+0.04}_{-0.04}$ \\
CXOXB J142435.8+325131 &  14 24 35.86 & 32 51 31.40 & 2.91 & 7 & 5&2&1.71$\pm$0.80 & 0.74$\pm$0.42 & 0.94$\pm$1.14&4708.48&0.868&  \\
CXOXB J142437.6+331152 &  14 24 37.63 & 33 11 52.52 & 2.15 & 9 & 6&3&2.24$\pm$0.87 & 0.91$\pm$0.45 & 1.46$\pm$1.26&4714.68&0.854&  \\
CXOXB J142437.9+354404 &  14 24 37.91 & 35 44 04.42 & 1.79 & 10 & 8&2&2.20$\pm$0.84 & 1.06$\pm$0.46 & 0.84$\pm$1.07&5038.95&0.856&-0.63$^{
+0.15}_{-0.13}$ \\
CXOXB J142438.1+334245 &  14 24 38.19 & 33 42 45.36 & 2.86 & 4 & 2&2&0.93$\pm$0.67 & 0.28$\pm$0.33 & 0.93$\pm$1.14&4714.68&0.873&  \\
CXOXB J142438.5+322649 &  14 24 38.56 & 32 26 49.68 & 2.05 & 8 & 6&2&1.70$\pm$0.78 & 0.77$\pm$0.42 & 0.82$\pm$1.06&5048.13&0.878&  \\
CXOXB J142439.8+340757 &  14 24 39.86 & 34 07 57.76 & 1.94 & 10 & 7&3&2.58$\pm$0.90 & 1.09$\pm$0.47 & 1.52$\pm$1.26&4714.68&0.832&-0.43$^{
+0.14}_{-0.13}$ \\
CXOXB J142440.0+345137 &  14 24 40.09 & 34 51 37.35 & 2.32 & 7 & 7&0&1.78$\pm$0.79 & 1.08$\pm$0.47 & $\le$0.7&4714.68&0.844&  \\
CXOXB J142440.4+351921 &  14 24 40.48 & 35 19 21.00 & 2.28 & 4 & 3&1&0.97$\pm$0.67 & 0.44$\pm$0.37 & 0.45$\pm$1.00&4711.62&0.869&  \\
CXOXB J142441.1+342619 &  14 24 41.14 & 34 26 19.14 & 2.37 & 6 & 5&1&1.49$\pm$0.75 & 0.75$\pm$0.42 & 0.46$\pm$0.99&4714.68&0.862&  \\
CXOXB J142441.2+354237 &  14 24 41.21 & 35 42 37.21 & 1.68 & 7 & 5&2&1.56$\pm$0.74 & 0.67$\pm$0.39 & 0.88$\pm$1.06&5038.95&0.844&  \\
CXOXB J142441.3+342703 &  14 24 41.31 & 34 27 03.00 & 2.22 & 8 & 5&3&2.01$\pm$0.83 & 0.75$\pm$0.42 & 1.50$\pm$1.25&4714.68&0.860&  \\
CXOXB J142441.9+345553 &  14 24 41.96 & 34 55 53.63 & 1.59 & 10 & 6&4&2.43$\pm$0.90 & 0.87$\pm$0.45 & 1.95$\pm$1.35&4714.68&0.891&-0.21$^{
+0.14}_{-0.14}$ \\
CXOXB J142442.4+325407 &  14 24 42.46 & 32 54 07.68 & 1.68 & 6 & 5&1&1.46$\pm$0.76 & 0.74$\pm$0.42 & 0.44$\pm$1.00&4708.48&0.872&  \\
CXOXB J142442.5+340817 &  14 24 42.52 & 34 08 17.64 & 1.31 & 15 & 11&4&3.68$\pm$1.04 & 1.62$\pm$0.55 & 1.94$\pm$1.36&4714.68&0.884&-0.48$^
{+0.09}_{-0.09}$ \\
CXOXB J142442.6+344923 &  14 24 42.67 & 34 49 23.49 & 2.50 & 5 & 3&2&1.23$\pm$0.71 & 0.45$\pm$0.37 & 0.97$\pm$1.14&4714.68&0.859&  \\
CXOXB J142442.8+333532 &  14 24 42.85 & 33 35 32.81 & 0.87 & 31 & 26&5&7.77$\pm$1.39 & 3.90$\pm$0.77 & 2.48$\pm$1.45&4714.68&0.873&-0.69$^
{+0.05}_{-0.04}$ \\
CXOXB J142442.9+325551 &  14 24 42.90 & 32 55 51.65 & 2.94 & 7 & 5&2&1.72$\pm$0.80 & 0.75$\pm$0.42 & 0.94$\pm$1.14&4708.48&0.858&  \\
CXOXB J142443.7+342538 &  14 24 43.78 & 34 25 38.31 & 1.22 & 12 & 9&3&2.94$\pm$0.96 & 1.32$\pm$0.51 & 1.45$\pm$1.25&4714.68&0.888&-0.51$^{
+0.12}_{-0.11}$ \\
CXOXB J142443.8+354143 &  14 24 43.83 & 35 41 43.02 & 1.79 & 5 & 4&1&1.14$\pm$0.67 & 0.55$\pm$0.37 & 0.44$\pm$0.93&5038.95&0.826&  \\
CXOXB J142443.8+322526 &  14 24 43.87 & 32 25 26.56 & 0.90 & 17 & 7&10&3.72$\pm$1.02 & 0.91$\pm$0.44 & 4.44$\pm$1.69&5048.13&0.872&0.18$^{
+0.08}_{-0.08}$ \\
CXOXB J142444.9+345552 &  14 24 44.97 & 34 55 52.60 & 1.98 & 5 & 4&1&1.23$\pm$0.71 & 0.60$\pm$0.40 & 0.46$\pm$0.99&4714.68&0.867&  \\
CXOXB J142445.3+323004 &  14 24 45.36 & 32 30 04.83 & 1.54 & 14 & 9&5&2.97$\pm$0.94 & 1.14$\pm$0.48 & 2.13$\pm$1.35&5048.13&0.895&-0.29$^{
+0.10}_{-0.10}$ \\
CXOXB J142445.5+331437 &  14 24 45.51 & 33 14 37.72 & 0.52 & 27 & 4&23&6.55$\pm$1.31 & 0.57$\pm$0.39 & 11.34$\pm$2.49&4714.68&0.906&0.71$^
{+0.05}_{-0.05}$ \\
CXOXB J142445.8+342945 &  14 24 45.87 & 34 29 45.37 & 1.67 & 4 & 4&0&0.96$\pm$0.67 & 0.58$\pm$0.39 & $\le$0.7&4714.68&0.893&  \\
CXOXB J142446.2+334013 &  14 24 46.21 & 33 40 13.64 & 1.32 & 7 & 6&1&1.67$\pm$0.79 & 0.86$\pm$0.45 & 0.46$\pm$0.99&4714.68&0.911&  \\
CXOXB J142446.3+345334 &  14 24 46.37 & 34 53 34.52 & 1.57 & 4 & 4&0&0.94$\pm$0.67 & 0.57$\pm$0.39 & $\le$0.7&4714.68&0.914&  \\
CXOXB J142447.6+334459 &  14 24 47.67 & 33 44 59.17 & 2.52 & 6 & 5&1&2.67$\pm$0.76 & 1.34$\pm$0.42 & 0.85$\pm$1.00&4714.68&0.484&  \\
CXOXB J142447.9+324935 &  14 24 47.93 & 32 49 35.48 & 0.60 & 28 & 20&8&6.76$\pm$1.33 & 2.88$\pm$0.69 & 3.91$\pm$1.68&4708.48&0.913&-0.43$^
{+0.05}_{-0.05}$ \\
\hline
\caption{ Properties of Point Sources in the XBo{\"o}tes Survey. 
The hardness ratio (HR) is provided only for sources with $n\geq 10$
    counts with uncertainties that are 90\% confidence limits derived
    using the maximum likelihood method of \citet{DHarris1993}.  The
    uncertainty in the source position is estimated from the count rate
    and the 50\% encircled energy radius at the source position relative
    to the pointing center (see text).  The complete table is presented
    only in the electronic edition of the Journal.
}
\label{tab:tab1}
\end{longtable}
\end{landscape}
\normalsize


\end{document}